\documentclass[preprint,superscriptaddress,amsmath,amssymb,prd,aps,showpacs,floatfix,nofootinbib]{revtex4-1}
\usepackage{graphicx}
\usepackage{dcolumn}
\usepackage{bm}
\usepackage{mathrsfs}
\usepackage{hyperref}
\usepackage{amsmath}
\usepackage{amssymb}
\usepackage{bm}
\usepackage{color}
\usepackage{float}
\usepackage{dcolumn}
\usepackage{multirow}
\usepackage{changepage}
\usepackage{enumerate}
\usepackage[normalem]{ulem}
\usepackage[dvipsnames]{xcolor}
\usepackage{braket}
\usepackage{nicefrac}
\usepackage{multirow}
\usepackage{tabularx}
 \usepackage{appendix}

\hypersetup{pdftex,colorlinks=true,linkcolor=blue,citecolor=red,menucolor=black,urlcolor=blue,filecolor=blue}

\newcommand{\mev}{\textrm{ MeV}}

\begin{document}
\title{The $D_s^+ \to a_0(980) e^+ \nu_e$ reaction and the $a_0(980)-f_0(980)$ mixing}
\date{\today}

\author{Jia-Xin Lin}
\affiliation{Department of Physics, Guangxi Normal University, Guilin 541004, China}

\author{Jia-Ting Li}
\affiliation{Department of Physics, Guangxi Normal University, Guilin 541004, China}

\author{Sheng-Juan Jiang}
\affiliation{Department of Physics, Guangxi Normal University, Guilin 541004, China}

\author{Wei-Hong Liang}
\email{liangwh@gxnu.edu.cn}
\affiliation{Department of Physics, Guangxi Normal University, Guilin 541004, China}
\affiliation{Guangxi Key Laboratory of Nuclear Physics and Technology, Guangxi Normal University, Guilin 541004, China}

\author{E. Oset}
\email{oset@ific.uv.es}
\affiliation{Department of Physics, Guangxi Normal University, Guilin 541004, China}
\affiliation{Departamento de F\'{\i}sica Te\'orica and IFIC, Centro Mixto Universidad de
Valencia-CSIC Institutos de Investigaci\'on de Paterna, Aptdo.22085,
46071 Valencia, Spain}


\begin{abstract}
  We perform a study of the $D_s^+ \to a_0(980)(f_0(980)) e^+ \nu_e$ reactions
  investigating the different sources of isospin violation
  which make the production of the $a_0(980)$ possible.
  We find that loops involving kaons in the production mechanism provide a source of isospin violation
  since they do not cancel due to the different mass of charged and neutral kaons,
  but we also find that the main source comes from the breaking of isospin in the meson-meson transition $T$ matrices,
  which contain information on the nature of the low lying scalar mesons.
  The reaction is thus very sensitive to the nature of the $a_0(980)$ and $f_0(980)$ resonances.
  Our results are consistent with the present upper bound for $a_0(980)$ production and only a factor three smaller,
  indicating that future runs with more statistics should find actual numbers for this reaction
  from where we can learn more about the origin of the scalar resonances and their nature.
\end{abstract}

\maketitle

\section{Introduction}
\label{sec:intro}

The nature of the low lying scalar mesons has been a subject of continuous debate \cite{1us,2us,3us,4us,5us,6us,7us}.
While the $a_0(980)$ was early advocated as a $K \bar K $ molecule \cite{14us},
the consideration of chiral dynamics for the interaction of mesons by means of chiral Lagrangians \cite{8us,9us},
and the subsequent unitarization with coupled channels introduced with the chiral approach \cite{10us,11us,12us,13us},
allows one to make this idea more quantitative
and it is found that the $f_0(500), f_0(980), a_0(980)$ resonances are dynamically generated from the interaction of pairs of mesons,
$\pi \pi, K \bar K, \eta \eta, \pi \eta$ with the dynamics dictated by the chiral Lagrangians.
A recent recollection of reactions and studies giving support to this picture can be found in the recent paper \cite{toledoliang}.

    One of the issues which has been advocated as a way to find information on the nature of the low lying scalar mesons
    is the one of the $a_0(980)-f_0(980)$ mixing.
    This occurs in reactions where the production of one of the states is allowed by isospin conservation
    and the other meson is observed in an isospin violating mode.
    The topic has been studied in numerous works \cite{2wang,3wang,4wang,5wang,6wang,7wang,8wang,9wang,10wang,
    11wang,12wang,13wang,14wang,15wang,16wang,17wang,18wang,19wang,20wang,
    21wang,22wang,23wang,24wang,25wang,vinibayar,aliev,chengli,liangchen,chengyu}
    and the relevant ingredient in the mixing lies in the mass difference between the neutral and charged kaons \cite{2wang,6wang,8wang,14wang,17wang,19wang,20wang,21wang,22wang,24wang,25wang}
    which is responsible for the lack of cancellation in loops involving kaons.
    This idea was first discussed in Ref. \cite{2wang}.

   While in most works one aims at determining the mixing in terms of a universal mixing parameter,
   it has been recently stressed that this concept is not adequate to deal with the mixing
   because it depends very much on the particular reaction and the mechanisms involved in it.
   Actually, the differences between the mixing observed in different reactions do indeed offer valuable information
   about the nature of the resonances.
   One surprise in this direction was shown
   by the abnormal isospin breaking in the decay $\eta(1405) \to \pi^0 f_0(980)$
   compared to the isospin allowed $\eta(1405) \to \pi^0 a_0(980)$ decay observed by the BES collaboration in Ref. \cite{beseta}.
   This was explained in Refs.~\cite{20wang,21wang,wuwu,achaeta,duzhao} by means of a triangle mechanism for the decay
   which develops a triangle singularity \cite{karplus,landau}.
   The idea of enhancing the isospin violation,
   and hence the mixing of $a_0(980)-f_0(980)$ has been exploited later
   and several reactions have been proposed: in Ref.~\cite{abnormal} an abnormal isospin violation
   and $a_0 -f_0$ mixing were studied in the $D_s^+ \to \pi^+ \pi^0 a_0(980) (f_0(980))$ reactions;
   In Ref.~\cite{isoviobdec} a triangle singularity is shown
   to enhance the isospin forbidden decay  $\bar B_s^0 \to J/\psi \pi^0 f_0(980)$
   versus the corresponding $a_0(980)$ production;
   In Ref.~\cite{daits} a triangle singularity in the $\tau^- \to \nu_\tau \pi^- f_0(980)$ ($a_0(980)$) decays
   is also responsible for an abnormal $a_0(980)-f_0(980)$ mixing.
   Similar results are found in the study of the ${J/\psi\rightarrow\eta\pi^0\phi}$ and ${\pi^0\pi^0\phi}$ decays,
   where a triangle singularity enhances again the isospin violation and the mixing of $a_0(980)-f_0(980)$ \cite{fengkunsakai}.

  In the present work we wish to study, yet,
  another reaction providing an example of $a_0(980)-f_0(980)$ mixing,
  taking advantage of a recent BESIII experiment \cite{besexp}  that studies the  $D_s^+\to a_0(980)^0e^+\nu_e$ decay.
  The $a_0(980)$ production in this reaction violates isospin in the Cabibbo favored mode,
  unlike the $f_0(980)$ production, which has been measured and is reported in the PDG \cite{pdg}.
  The $a_0(980)$ is not identified in the reaction \cite{besexp},
  and only one upper limit for the branching ratio of the reaction is provided. On the other hand,
  there is already a  theoretical work devoted to this reaction in Ref.~\cite{weiwang},
  where combining theoretical amplitudes in semileptonic decay,
  together with experimental information and magnitudes evaluated in other works,
  rates for the ratio of $a_0(980)$ to $f_0(980)$ production are obtained, ranging around 1\%.
  In view of the new experiment and the likely precise measurement of the rates for the reaction with improved statistics in the future,
  we present here a detailed and accurate calculation of this rate
  based upon the picture of the
 $a_0(980)$ and the $f_0(980)$ as dynamically generated resonances within the chiral unitary approach.
 The picture can relate the $f_0(980)$ and $a_0(980)$ production rates
 and then we rely upon the measurements for the $D_s^+\to f_0(980)^0e^+\nu_e$ reaction
 to obtain information on the $D_s^+\to a_0(980)^0e^+\nu_e$ decay,
 which can be contrasted with future measurements.
 The picture follows closely the developments on the production of dynamically generated resonances
 in weak decays of hadrons described in Ref.~\cite{review}.
 In particular, the   $D_s^+\to f_0(980)^0e^+\nu_e$ decay is studied in detail in Ref.~\cite{sekisemi},
 together with other related reactions.
 We will present results on the $\pi^0 \eta$ mass distribution in $D_s^+\to a_0(980)^0e^+\nu_e; a_0 \to \pi^0 \eta$
 versus the $\pi^+ \pi^-$ mass distribution  in $D_s^+\to f_0(980)^0e^+\nu_e; f_0 \to \pi^+ \pi^-$,
 and will show that the ratio obtained for the integrated mass distributions is consistent with
 the present upper bound for the reaction \cite{besexp} and not too far, thus,
 giving  incentives for an improved measurement with more statistics,
 which, together with other cases already studied in the literature
 will provide good information concerning the nature of the low lying scalar mesons.

\section{Formalism}
\label{sec:form}

\begin{figure}[t]
   \includegraphics[width=0.4\linewidth]{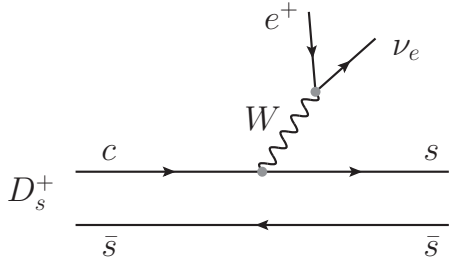}
   \vspace{-0.4cm}
    \caption{Semileptonic $D_s^+$ decay at the quark level. }
    \label{fig:quarklev}
\end{figure}

We follow the formalism developed in Ref.~\cite{sekinavarra}, particularized to $D_s$ semileptonic decay in Ref.~\cite{sekisemi}.
The approach has also been applied to the study of semileptonic decay of baryons in Refs.~\cite{liangxie} and \cite{pavaoliang}.
We start from the semileptonic $D_s^+$ decay at the quark level depicted in Fig.~\ref{fig:quarklev}.
We observe that one gets an $s\bar s$ pair at the end, which has isospin $I=0$.
This is the dominant Cabibbo decay mode.
Note that the Cabibbo suppressed mode would produce a $d\bar s$ state, with $I=\frac{1}{2}$,
which cannot lead to $f_0$ or $a_0$ production in $I=0, 1$ respectively.
With the topology of Fig.~\ref{fig:quarklev} we can obtain the $f_0(980)$ state with isospin conservation,
but the production of $a_0(980)$ is isospin forbidden and involves isospin violation.

The next step involves hadronization of the $s\bar s$ component introducing a $\bar q q$ pair with the vacuum quantum numbers.
We just need the isoscalar $\bar q q$ component, $\sum_i \bar q_i q ~(q_i=u, d, s)$.
Hence, the $s\bar s$ component becomes
\begin{equation}\label{eq:H1}
  s\bar s \to H = \sum_i s\, \bar q_i q \, \bar s = \sum_i {\mathcal{P}}_{3i} \; {\mathcal{P}}_{i3}\, = ({\mathcal{P}}^2)_{33},
\end{equation}
where ${\mathcal{P}}$ is the $q\bar q$ matrix in SU(3),
which we write in terms of mesons, in the present case, pseudoscalar mesons ($P$), to generate the $f_0$ and $a_0$ resonances.
We write the standard  ${\mathcal{P}}$ matrix, used in chiral perturbation theory \cite{9us},
where the $\eta$ is assumed to be $\eta_8$ of SU(3), as
\begin{equation}\label{eq:PNomix}
{\mathcal{P}} = \left(
           \begin{array}{ccc}
             \frac{1}{\sqrt{2}}\pi^0 + \frac{1}{\sqrt{6}}\eta + \frac{1}{\sqrt{3}}\eta' & \pi^+ & K^+ \\[2mm]
             \pi^- & -\frac{1}{\sqrt{2}}\pi^0 + \frac{1}{\sqrt{6}}\eta + \frac{1}{\sqrt{3}}\eta' & K^0 \\[2mm]
            K^- & \bar{K}^0 & -\sqrt{\frac{2}{3}}\eta + \sqrt{\frac{1}{3}}\eta' \\
           \end{array}
         \right).
\end{equation}
On the other hand, when the $\eta$ and $\eta'$ mixing is considered (we take the mixing of Ref.~\cite{bramon}) we have
\begin{equation}\label{eq:Pmix}
{\mathcal{P}}^{(\mathrm{m})} =
         \left(
           \begin{array}{ccc}
             \frac{1}{\sqrt{2}}\pi^0 + \frac{1}{\sqrt{3}}\eta + \frac{1}{\sqrt{6}}\eta' & \pi^+ & K^+ \\[2mm]
             \pi^- & -\frac{1}{\sqrt{2}}\pi^0 + \frac{1}{\sqrt{3}}\eta + \frac{1}{\sqrt{6}}\eta' & K^0 \\[2mm]
            K^- & \bar{K}^0 & -\frac{1}{\sqrt{3}}\eta + \sqrt{\frac{2}{3}}\eta' \\
           \end{array}
         \right).
\end{equation}
We ignore the $\eta'$ in our calculations since it plays no role in the building of the $f_0(980),\, a_0(980)$ resonances \cite{10us}.

The hadron component $H$ of Eq.~\eqref{eq:H1} is then given by
\begin{equation}\label{eq:HNomix}
  H= K^-K^+ + \bar K^0 K^0 + \frac{2}{3}\, \eta \eta
\end{equation}
when using matrix ${\mathcal{P}}$ of Eq.~\eqref{eq:PNomix} and
\begin{equation}\label{eq:Hmix}
  H= K^-K^+ + \bar K^0 K^0 + \frac{1}{3}\, \eta \eta
\end{equation}
when using matrix ${\mathcal{P}}^{(\mathrm{m})}$ of Eq.~\eqref{eq:Pmix}.
Only the $\eta \eta$ component is changed which affects the $f_0$ production but not the one of $a_0$.

\begin{figure}[b]
   \includegraphics[width=1\linewidth]{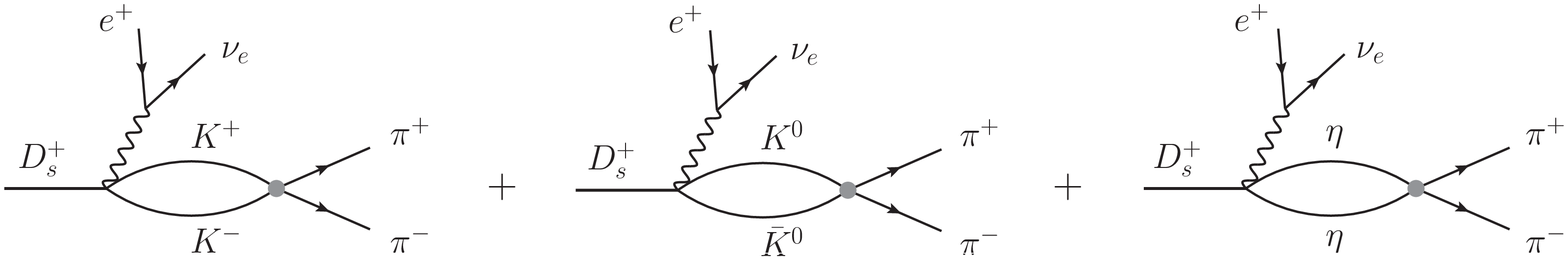}\\[0.2cm]
   \includegraphics[width=1\linewidth]{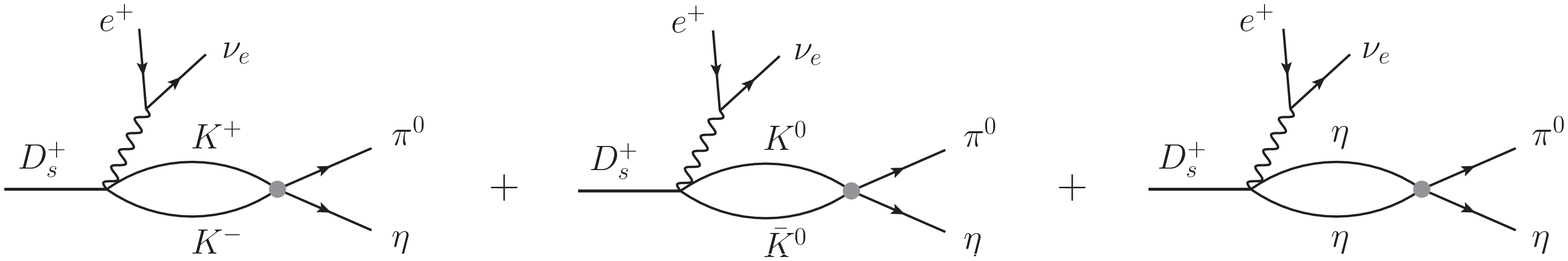}
    \caption{Final state interaction of the hadron components leading to $\pi^+\pi^-$ or $\pi^0 \eta$ in the final state.}
    \label{fig:FeynDiag}
\end{figure}

The next step consist of allowing the meson-meson components to undergo final state interaction,
which is depicted in Fig.~\ref{fig:FeynDiag}.
For this purpose we shall use the chiral unitary approach of Ref.~\cite{10us}
which produces very accurate meson-meson amplitudes from threshold to about $1200 \mev$ and generates the $f_0(980)$ and $a_0(980)$ resonances.
Note that the $\nu_e e^+$ system carries much of the energy of the $D_s$
and we are only concerned about the invariant mass of the $\pi^+\pi^-$ and $\pi^0\eta$ systems in a very narrow region of invariant mass around the peaks of the $f_0(980)$ and $a_0(980)$ resonances.

Note that neither Eq.~\eqref{eq:HNomix} nor Eq.~\eqref{eq:Hmix} contain $\pi^+\pi^-$ or $\pi^0 \eta$,
but the final state interaction of the $K\bar K$ and $\eta\eta$ components can lead to both.
In Fig.~\ref{fig:FeynDiag},
the circle following the meson-meson loop signifies the transition matrix from the $PP$ state to $\pi^+\pi^-$ or $\pi^0 \eta$,
which contains the $f_0(980)$ and $a_0(980)$ information respectively,
since in the approach of Ref.~\cite{10us}
that we follow these resonances are originated as a consequence of the $PP$ interaction in coupled channels.
We follow the chiral unitary approach of Ref.~\cite{10us} and write in matrix form the transition matrix $T$ as
\begin{equation}\label{eq:BSeq}
  T=[1-V\,G]^{-1}\, V,
\end{equation}
where $V$ is the transition potential between the coupled channels
and $G$ is the $PP$ loop function which we regularize by means of a cut-off, $q_{\rm max}$, in the three momentum of the loop.

The $G$ function in the cut-off regularization is given by \cite{10us}
\begin{equation}\label{eq:Gfuction}
  G(\sqrt{s})=\int_0^{q_{\rm max}} \frac{q^2\; {\rm d}q}{(2\pi)^2}\; \frac{w_1+w_2}{w_1\, w_2 \,[s-(w_1 +w_2)^2+i \epsilon]},
\end{equation}
with $w_i=\sqrt{m_i^2+\vec q^{\,2}}$ and $\sqrt{s}$ the centre-of-mass energy of the two mesons in the loop,
and $q_{\rm max}$ stands for the maximum value of the modulus of the three momentum $\vec q$ allowed in the integral of Eq.~\eqref{eq:Gfuction}.
The cut-off is needed to regularize the logarithmic divergence of Eq.~\eqref{eq:Gfuction}.
The value of $q_{\rm max}$ should have a natural value of $500 \mev - 1000 \mev$
and the precise value of it is chosen to fit one particular experimental magnitude.
In Refs.~\cite{LiangOset,DaiXie}, which consider the $\pi\pi, \pi\eta, \eta\eta, K\bar K$ channels,
a cut-off $q_{\rm max}=600 \mev$ was used to reproduce the $f_0(980)$ mass and then other magnitudes were fairly well reproduced.

We calculate $V$ both with the ${\mathcal{P}}$ and ${\mathcal{P}}^{(\mathrm{m})}$ matrices
and show the matrix elements of $V$ in Appendix A for the $s$-wave that we consider.

Defining the weight of the $PP$ components in $H$ as
\begin{equation}\label{eq:weight}
  h_{K^+K^-}=1,~~~~ h_{K^0 \bar K^0}=1,~~~~ h_{\eta\eta}=\frac{2}{3},~~~~ h_{\eta\eta}^{(\mathrm{m})}=\frac{1}{3},
\end{equation}
we can write the amplitude for the $D_s^+\to e^+\nu_e \pi^+\pi^-$ decay as
\begin{eqnarray}\label{eq:tpipi}
t_{\pi^+\pi^-} &=& {\mathcal{C}}\left[ h_{K^+K^-} \cdot G_{K^+K^-} (M_{\rm inv}(\pi^+\pi^-)) \cdot T_{K^+K^-,\pi^+\pi^-}(M_{\rm inv}(\pi^+\pi^-))
\right.\nonumber \\
&&~ + h_{K^0 \bar K^0} \cdot G_{K^0 \bar K^0} (M_{\rm inv}(\pi^+\pi^-)) \cdot T_{K^0 \bar K^0,\pi^+\pi^-}(M_{\rm inv}(\pi^+\pi^-))\nonumber \\
&&~ + h_{\eta\eta} \times 2\times \frac{1}{2} \cdot \left. G_{\eta\eta} (M_{\rm inv}(\pi^+\pi^-)) \cdot T_{\eta\eta,\pi^+\pi^-}(M_{\rm inv}(\pi^+\pi^-))\right],
\end{eqnarray}
where, ${\mathcal{C}}$ is an arbitrary normalization constant which will be canceled in the ratio of the $a_0(980), f_0(980)$ production rates,
$T_{i,j}$ is the total amplitude for the $i\to j$ transition
which can be obtained by solving Eq.~\eqref{eq:BSeq}, the Bethe-Salpeter equation in coupled channel.
The factor $2$ accompanying $G_{\eta\eta}$ stands for the two combinations to produce $\eta\eta$ from $H$,
and the factor $\frac{1}{2}$ for the identity of the $\eta\eta$ particles in the loop.
We use the unitary normalization to evaluate Eq.~\eqref{eq:tpipi} with the $\eta\eta$ state being normalized to $\frac{1}{\sqrt 2} \eta\eta$,
and at the end we must write $T_{\eta\eta,\pi^+\pi^-}= \sqrt{2} T^{(u)}_{\eta\eta,\pi^+\pi^-}$ (with $\eta\eta$ unitary normalization),
and the same for $T_{\eta\eta,\pi^0\eta}$.

Similarly we have the amplitude for the $D_s^+\to e^+\nu_e \pi^0\eta$ decay
\begin{eqnarray}\label{eq:tpieta}
t_{\pi^0\eta} &=& {\mathcal{C}}\left[ h_{K^+K^-} \cdot G_{K^+K^-} (M_{\rm inv}(\pi^0\eta)) \cdot T_{K^+K^-,\pi^0\eta}(M_{\rm inv}(\pi^0\eta))
\right.\nonumber \\
&&~ + h_{K^0 \bar K^0} \cdot G_{K^0 \bar K^0} (M_{\rm inv}(\pi^0\eta)) \cdot T_{K^0 \bar K^0,\pi^0\eta}(M_{\rm inv}(\pi^0\eta))\nonumber \\
&&~ + h_{\eta\eta} \times 2\times \frac{1}{2} \cdot \left. G_{\eta\eta} (M_{\rm inv}(\pi^0\eta)) \cdot T_{\eta\eta,\pi^0\eta}(M_{\rm inv}(\pi^0\eta))\right].
\end{eqnarray}
We would have the corresponding expressions with the $\eta-\eta'$ mixing
by changing $h_{\eta\eta}$ to $h_{\eta\eta}^{(\mathrm{m})}$ in Eqs.~\eqref{eq:tpipi} and \eqref{eq:tpieta}.

Expressions for the double differential decay width ${\mathrm d}^2 \Gamma / {\mathrm d} M_{\rm inv}(\nu l)\,{\mathrm d} M_{\rm inv}(ij)$
are given explicitly in Refs.~\cite{sekinavarra,sekisemi},
by means of which we immediately write the differential decay width
\begin{eqnarray}\label{eq:dGam}
\frac{{\mathrm d} \Gamma}{{\mathrm d} M_{\rm inv}(ij)}
&=& \frac{|G_F\, V_{cs}|^2}{32\, \pi^5\, m_{D_s}^3 }\, \frac{1}{M_{\rm inv}(ij)} \;|t_{ij}|^2 \nonumber \\
&&  \times \int d M_{\mathrm{inv}}(\nu l) \cdot P_{\mathrm{cm}} \cdot
		\tilde{p}_{i} \cdot \tilde{p}_{\nu} \cdot \left[M_{\mathrm{inv}}(\nu l)\right]^2 \cdot
        \left(\tilde{E}_{D_s}\, \tilde{E}_{ij}-\frac{1}{3}\,|\tilde{p}_{D_s}|^2 \right),
\end{eqnarray}
where, $M_{\rm inv}(ij)$ is the invariant mass of the final $\pi^+\pi^-$ or $\pi^0\eta$,
$M_{\rm inv}(\nu l)$ is the invariant mass of the $\nu l$ pair,
taking values in the range of $[m_e+m_\nu,\, M_{D_s}-M_{\rm inv}(ij)]$,
$V_{cs}$ is the Cabibbo-Kobayashi-Maskawa (CKM) matrix element and $G_F$ the Fermi coupling constant $G_F=g_W^2/(4\sqrt{2}\, M_W^2)$,
with $M_W$ the mass of the $W$ boson and $g_W$ the weak coupling constant,
\begin{equation}\label{eq:GF}
  G_F= 1.166\times 10^{-5}\; {\rm GeV}^{-2}.
\end{equation}
In Eq.~\eqref{eq:dGam}, $P_{\mathrm{cm}}$ is the momentum of the $(\nu l)$ system in the $D_s$ rest frame,
$\tilde{p}_{i}$ is the momentum of the $i$ meson in the $ij$ rest frame,
$\tilde{p}_{\nu}$ is the neutrino momentum in the $\nu l$ rest frame,
$\tilde{E}_{D_s}$ is the energy of the $D_s$ in the $\nu l$ rest frame,
$\tilde{E}_{ij}$ is the energy of the $ij$ pair of pseudoscalar in the $\nu l$ rest frame,
and $\tilde{p}_{D_s}$ is the momentum of the $D_s$ in the $\nu l$ rest frame,
\begin{equation}
P_{\mathrm{cm}}=\frac{\lambda^{1/2}(M_{D_s}^{2}, \, M^2_{\mathrm{inv}}(\nu l), \,M_{\mathrm{inv}}^2(ij))}{2 M_{D_s}},
\end{equation}
\begin{equation}
\tilde{p}_{i}=\frac{\lambda^{1/2}(M_{\mathrm{inv}}^2(ij), \,m_{i}^{2}, \, m_{j}^{2})}{2 M_{\mathrm{inv}}{(ij)}},
\end{equation}
\begin{equation}
\tilde{p}_{\nu}=\frac{\lambda^{1/2}(M^2_{\mathrm{inv}}(\nu l), \, m_{\nu}^{2}, \, m_{l}^{2})}{2 M_{\mathrm{inv}}{(\nu l)}},
\end{equation}
\begin{equation}
\tilde{E}_{D_s}=\frac{M_{D_s}^{2} + M^2_{\mathrm{inv}}(\nu l) - M^2_{\mathrm{inv}}(ij)}{2 M_{\mathrm{inv}}(\nu l)},
\end{equation}
\begin{equation}
\tilde{E}_{ij}=\frac{M_{D_s}^{2} - M^2_{\mathrm{inv}}(\nu l) - M^2_{\mathrm{inv}}(ij)}{2 M_{\mathrm{inv}}(\nu l)},
\end{equation}
\begin{equation}
\tilde{p}_{D_s}=\sqrt{\tilde{E}_{D_s}^2-M^2_{D_s}}.
\end{equation}
Note that $\tilde{E}_{D_s}-\tilde{E}_{ij}=M_{\mathrm{inv}}{(\nu l)}$ as it should be.

Note that $t_{ij}$ entering Eq.~\eqref{eq:dGam} has a constant ${\mathcal{C}}$ shown in Eqs.~\eqref{eq:tpipi} and \eqref{eq:tpieta}.
This corresponds to matrix elements involving quark wave functions and the process of hadronization,
which we do not evaluate here.
This is also not evaluated in Ref.~\cite{sekinavarra} but rather it is taken by fitting some observed semileptonic transition,
and values around ${\mathcal{C}}\simeq 7$ are obtained.
The interesting thing of our approach is that we can compare one semileptonic decay to another related one,
involving the same elementary process, where the constant ${\mathcal{C}}$ cancels in the ratio,
and based upon one of them we can calculate rates for the other one and mass distributions.
This is what we shall do here comparing the $a_0(980)$ production with the $f_0(980)$ one,
which is observed experimentally.
The method avoids theoretical uncertainties of the microscopical models of the reactions
and provides a very accurate tool to make predictions for one reaction based on experimental information of a related one.

\section{Results}
\label{sec:result}
In the first place we show in Table \ref{tab:poleNomix} the results for the poles of the scalar resonances
obtained for different values of $q_{\rm max}$ for the $f_0(500)$ (first line), the $f_0(980)$ (second line) and $a_0(980)$ (third line),
using the matrix ${\mathcal{P}}$ of Eq.~\eqref{eq:PNomix}.
In Table \ref{tab:poleMix} we show the equivalent results for the case
when we use the matrix ${\mathcal{P}}^{(\mathrm{m})}$ of Eq.~\eqref{eq:Pmix} accounting for the $\eta-\eta'$ mixing.
\begin{table}[h]
\renewcommand\arraystretch{1.0}
\centering
\caption{\vadjust{\vspace{-2pt}}Poles for $J^{PC}=0^{-+}$ $PP$ states for the case without $\eta-\eta'$ mixing. [in MeV]}\label{tab:poleNomix}
\begin{tabular*}{1.00\textwidth}{@{\extracolsep{\fill}}cccc}
\hline\hline
 $q_{\rm max}$    &$600$               &$800$               & $1020$            \\
\hline
                  &$452.91+i252.92$    & $466.26 +i219.28$  & $469.58+i187.92$  \\
                  &$981.25+i5.40$      &  $940.99 +i13.87$  & $892.62+i19.71$   \\
                  &                    &                    & $985.24+i57.78$        \\
\hline
\hline
\end{tabular*}
\end{table}
\begin{table}[h]
\renewcommand\arraystretch{1.0}
\centering
\caption{\vadjust{\vspace{-2pt}}Poles for $J^{PC}=0^{-+}$ $PP$ states for the case with standard $\eta-\eta'$ mixing. [in MeV]}\label{tab:poleMix}
\begin{tabular*}{1.00\textwidth}{@{\extracolsep{\fill}}cccc}
\hline\hline
 $q_{\rm max}$    & $650$                       & $750$             & $1000$            \\
\hline
                  & $457.54+i244.25$            & $464.11+i227.33$  & $469.59+i190.38$  \\
                  & $985.45+i6.05$              & $972.93+i11.42$   & $933.17+i20.72$   \\
                  & $(m_{K^0\bar K^0}=995.2)$   & $994.06+i38.70$   & $984.34+i63.74$        \\
\hline
\hline
\end{tabular*}
\end{table}
We can see that for $q_{\rm max} =600 \mev$ we obtain values of the masses of the $f_0(500)$ and $f_0(980)$ in fair agreement with experiment.
The $a_0(980)$ does not show up with this value of $q_{\rm max}$,
but it shows up as a cusp at the $K\bar K$ threshold \cite{liangchic1},
which is in very good agreement with the high precision experiment on $a_0(980)$ production in Ref.~\cite{kornicer}.
We, thus, take $q_{\rm max} =600 \mev$ for our calculations as done in Ref.~\cite{liangchic1} 
for the case without $\eta -\eta'$ mixing and $q_{\rm max} = 650 \mev$ for the case with $\eta -\eta'$ mixing.

\begin{figure}[b]
\begin{center}
\includegraphics[scale=0.75]{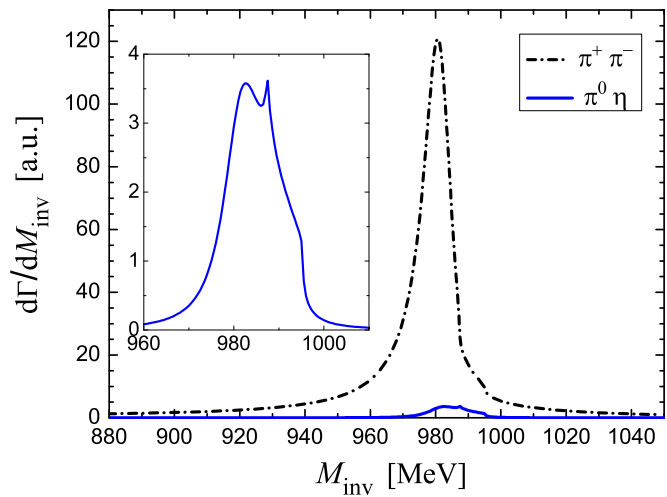}
\end{center}
\vspace{-1cm}
\caption{$M_{\rm inv}(\pi^+ \pi^-)$ mass distribution for $D_s^+ \to e^+ \nu_e f_0(980), f_0(980) \to \pi^+ \pi^-$ decay, and $M_{\rm inv}(\pi^0 \eta)$ mass distribution for $D_s^+ \to e^+ \nu_e a_0(980), a_0(980) \to \pi^0 \eta$ decay. (Without $\eta-\eta'$ mixing)}
\label{Fig:MinvNoMix}
\end{figure}

\begin{figure}[t]
\begin{center}
\includegraphics[scale=0.75]{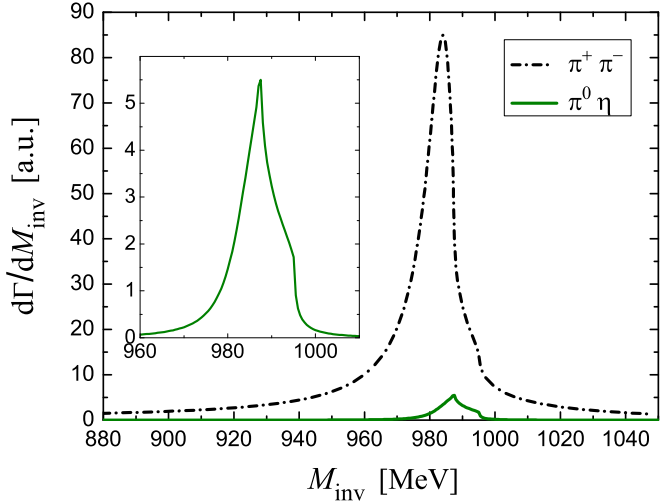}
\end{center}
\vspace{-1cm}
\caption{$M_{\rm inv}(\pi^+ \pi^-)$ mass distribution for $D_s^+ \to e^+ \nu_e f_0(980), f_0(980) \to \pi^+ \pi^-$ decay, and $M_{\rm inv}(\pi^0 \eta)$ mass distribution for $D_s^+ \to e^+ \nu_e a_0(980), a_0(980) \to \pi^0 \eta$ decay. (With $\eta-\eta'$ mixing)}
\label{Fig:MinvMix}
\end{figure}

In Figs.~\ref{Fig:MinvNoMix} and \ref{Fig:MinvMix}
we show results for $\dfrac{{\mathrm d}\Gamma}{{\mathrm d} M_{\rm inv}(ij)}$ for $\pi^+\pi^-$ and $\pi^0\eta$
without and with $\eta-\eta'$ mixing, respectively.
The results are very similar, and the differences we take as an estimation of uncertainties of our formalism.
We observe that there is a neat signal for $f_0(980)$ production, but also a small contribution for the $a_0(980)$ production.
The shape of the $f_0(980)$ mass distribution is the standard one corresponding to an isospin allowed reaction.
The width reflects the ordinary width of the $f_0(980)$ for its decay into $\pi^+\pi^-,\, \pi^0\pi^0$.
The shape of the $a_0(980)$ production is very narrow, about $10-15 \mev$ width,
and quite distinct from the standard $a_0(980)$ cusp-like shape (see Refs.~\cite{kornicer,liangchic1}).
In our approach it is tied to the mass difference between the $K^+$ and the $K^0$,
this is about $4\mev$, very different from the apparent width of about $120 \mev$ in Ref.~\cite{kornicer}.
This comes from the interaction of pseudoscalar mesons
and what generates the resonance is the loop iteration of the $PP$ potential.
In our formalism,
if in the loops we use average masses for $K^+ (K^-)$ and $K^0 (\bar K^0)$ we obtain a zero strength for $a_0(980)$ production
and there is an exact cancellation of the $K^+ K^-$ and $K^0 \bar K^0$ loops in Fig.~\ref{fig:FeynDiag}.
Our picture is actually a bit different from other pictures where the $K^+,\, K^0$ mass difference is the source for the isospin violation.
In our case this difference is also the source of the violation,
but it comes in two ingredients in Fig.~\ref{fig:FeynDiag},
the explicit $K\bar K$ loop and the transition $T_{i,j}$ matrix.
Indeed in the generation of the $T$ matrix via Eq.~\eqref{eq:BSeq} we are using also the $G$ functions in coupled channels.
If we use average $K$ masses in the loop, the $T$ matrix conserves isospin,
while there is some isospin violation in the $T$ matrix when we use actual $K^+, K^0$ masses in the loops.

Upon integration of Eq.~\eqref{eq:dGam} over $M_{\rm inv}(ij)$, one obtains the decay width $\Gamma$. We define the ratio $R$ as
\begin{equation}\label{eq:R}
  R=\frac{\Gamma(D_s^+ \to e^+ \nu_e a_0(980), a_0(980) \to \pi^0 \eta )}
        {\Gamma(D_s^+ \to e^+ \nu_e f_0(980), f_0(980) \to \pi^+ \pi^-)},
\end{equation}
and show values of $R$ in Table \ref{tab:R} with different assumptions.
\begin{table*}[t]
\renewcommand\arraystretch{1.1}
\centering
\caption{\vadjust{\vspace{-2pt}}values of $R$ with different assumptions. (In the table, I.V. denotes isospin violation.)}\label{tab:R}
\begin{tabular}{l|l|c}
\hline \hline
\multirow{3}{*}{no $\eta-\eta'$ mixing}     & ~I.V. both in $T$ matrix and in explicit $K\bar K$ loops (Case 1) &$3.1\times 10^{-2}$\\
\cline{2-3}
                                            & ~I.V. only in $T$ matrix (Case 2)& ~~$3.5\times 10^{-2}$~~\\
\cline{2-3}
                                            & ~I.V. only in explicit $K\bar K$ loops (Case 3)& $7.1\times 10^{-4}$ \\
\hline\hline
\multirow{3}{*}{with $\eta-\eta'$ mixing}~  & ~I.V. both in $T$ matrix and in explict $K\bar K$ loops (Case 4)& $3.7\times 10^{-2}$ \\
\cline{2-3}
                                            &~I.V. only in $T$ matrix (Case 5) & $4.2\times 10^{-2}$\\
\cline{2-3}
                                            & ~I.V. only in explicit $K\bar K$ loops (Case 6)& $1.2\times 10^{-3}$ \\
\hline\hline
\end{tabular}
\end{table*}
 One can see that using the $\eta-\eta'$ mixing increases the ratio $R$ in about $20\%$.

 Inspection of Table~\ref{tab:R} shows interesting results.
 We observe that the main contribution to the width does not come from the explicit loops in Fig.~\ref{fig:FeynDiag}
 but from the isospin violation in the $T$ matrix of the chiral unitary approach.
 The consideration of the isospin violation in the $T$ matrix gives a rate about three times bigger
 than the contribution from the explicit loops in Fig.~\ref{fig:FeynDiag}.
 Hence, the assumption that the $f_0(980)$ and $a_0(980)$ resonances are dynamically generated from the meson-meson interaction
 has a sizable impact in the ratio of the $a_0(980)$ and $f_0(980)$ production rates in this decay mode.
 The  mode that we have studied is particularly suited to investigate the properties of these resonances because,
 unlike many other processes where $\pi^+ \pi^-$ and $\pi^0 \eta$ are produced at the tree level after the hadronization of quarks,
 in the present case these states are absent at the tree level and only the final state interaction,
 via the $T$ matrix that contains the information on the resonances, leads finally to these two pairs of mesons.
 It is thus a privileged case, where the decay amplitude is already directly proportional to the meson-meson transition matrices.

If we use the ratio $R$ of Table \ref{tab:R}
and the branching ratio of ${\mathrm{Br}}[D_s^+ \to e^+ \nu_e f_0(980), f_0(980) \to \pi^+ \pi^-]$ of the most recent experiment \cite{hietala},
also used in Ref.~\cite{besexp} for their analysis,
\begin{equation}\label{eq:Br}
  {\mathrm{Br}}[D_s^+ \to e^+ \nu_e f_0(980), f_0(980) \to \pi^+ \pi^-]=(0.13\pm 0.03\pm 0.01)\times 10^{-2},
\end{equation}
we obtain the branching ratio for $a_0(980)$ production
    \begin{align}\label{eq:Bra0}
        {\mathrm{Br}}[D_s^+ \to e^+ \nu_e a_0(980), a_0(980) \to \pi^0 \eta]=%
            \begin{cases}
                (4.0\pm 0.9)\times 10^{-5}, & {\rm for~ Case~1;} \\[0.1cm]
                (4.8\pm 1.1)\times 10^{-5}, & {\rm for~ Case~4.}
            \end{cases}
    \end{align}%
These results are compatible with the upper limit of the recent BESIII measurement \cite{besexp}
\begin{equation}\label{eq:BesBr}
  {\mathrm{Br}}[D_s^+ \to e^+ \nu_e a_0(980), a_0(980) \to \pi^0 \eta]< 1.2\times 10^{-4}.
\end{equation}
We can see that the results obtained are below the experimental upper bound, but not too far, a factor about three times smaller.
This is interesting to know in order to plan future searches.
The branching ratio and the shape of the $\pi^0\eta$ mass distributions are elements
that carry much information on the nature of the $f_0(980)$ and $a_0(980)$ resonances,
and its experimental determination will be very valuable.

\section{Conclusions}
\label{sec:concl}
We have studied in detail the $D_s^+ \to a_0(980)(f_0(980)) e^+ \nu_e$ reactions from the perspective
that the two resonances are dynamically generated from the interaction of pseudoscalar mesons.
We show that under strict isospin conservation, the production of the $a_0(980)$ is forbidden.
Yet, we find two sources of isospin violation that contribute to the production of the $a_0(980)$.
The isospin violation comes from the mechanism of production,
where an $s \bar s$ quark pair is originally produced which hadronizes in $K \bar K$ pairs and $\eta \eta$.
Interestingly the $\pi^+ \pi^-$ and the $\pi^0 \eta$,
which one needs to identify the $f_0(980)$ and $a_0(980)$ production, are not produced at that step.
Hence it is the rescattering of the $K \bar K, \eta \eta$ components that produce $\pi^+ \pi^-$ and the $\pi^0 \eta$ at the end.
The picture shows that the weak decay is proportional to the meson-meson transition $T$ matrix,
such that the information carried out by this magnitude on isospin violation shows up clearly.
But prior to the rescattering of these meson-meson components one has their propagation,
which is considered in terms of loop functions.
In these loop functions there is one source of isospin violation
since one finds that the loops containing $K^+ K^-$ or $K^0 \bar K^0$ do not cancel
due to the mass difference between the charged and neutral kaons.
However, we find also a source of isospin violation from the meson-meson transition matrices,
which are tied to the way the resonances are generated from the meson-meson interaction.
Actually we find that the latter contribution is about three times more important
than the one coming from the explicit loops in the weak decay.
This is telling us that this reaction is very sensitive to the way the resonances are generated and hence to their nature.

    When we compare to experiment we see that there is only one experiment, performed very recently at BESIII,
    which only provides an upper bound for the $a_0(980)$ production.
    Our results are consistent with this upper bound and are only about a factor three below the experimental threshold,
    indicating that future runs of the reaction with more statistics can already provide actual numbers.
    We also show that the shape of the $\pi^0 \eta$ mass distribution is very different to
    the one of the ordinary $a_0(980)$ production in isospin allowed reactions,
    which is tied once more to the difference of mass between the charged and neutral kaons.

   All together, our study clearly indicates that the observation of the $a_0(980)$ in this decay of the $D_s$
   provides very relevant information concerning the nature of the low lying scalar resonances
   and should give a motivation for its measuring with more statistics.

\begin{acknowledgments}
This work is  partly supported by the National Natural Science Foundation of China under Grants No. 11975083 and No. 12047567.
This work is also partly supported by the Spanish Ministerio de Economia y Competitividad
and European FEDER funds under Contracts No. FIS2017-84038-C2-1-P B
and by Generalitat Valenciana under contract PROMETEO/2020/023.
This project has received funding from the European Unions Horizon 2020 research and innovation programme
under grant agreement No. 824093 for the ``STRONG-2020" project.
\end{acknowledgments}

\appendix
\section{The $s$-wave $V_{ij}$ potentials with and without $\eta-\eta'$ mixing}
In this appendix we show the transition potentials $V_{ij}$ for the $PP \to PP$ scattering in $s$-wave,
which are used to generate the light scalar mesons $f_0(500), f_0(980)$ and $a_0(980)$
by solving the Bethe-Salpeter equation of Eq.~\eqref{eq:BSeq}.

Considering the quantum numbers of $f_0$ and $a_0$ mesons, we have six neutral $PP$ channels:
$\pi^+\pi^-$, $\pi^0\pi^0$, $K^+K^-$, $K^0\bar{K}^0$, $\eta\eta$ and $\pi^0\eta$, labeled by the indices $i=1, 2,...,6$ respectively.
The lowest order chiral Lagrangian for the pseudoscalar-pseudoscalar mesons interaction is given by \cite{10us}
\begin{equation}\label{eq:L}
	\mathcal{L}_2=\frac{1}{12f^2}\langle(\partial_\mu\Phi\Phi-\Phi\partial_\mu\Phi)^2+M\Phi^4\rangle,
\end{equation}
where, $f=93 \mev$ is the pion decay constant,
the symbol $\langle$ $\rangle$ stands for the trace of matrices,
$M$ is the pseudoscalar meson mass matrix given by
\begin{equation}
M=\left(\begin{array}{ccc}
m_{\pi}^{2} & 0 & 0 \\
0 & m_{\pi}^{2} & 0 \\
0 & 0 & 2 m_{K}^{2}-m_{\pi}^{2}
\end{array}\right).
\end{equation}
$\Phi$ is the SU(3) matrix of pseudoscalar mesons,
taking the form of ${\mathcal{P}}$ of Eq.~\eqref{eq:PNomix} without $\eta-\eta'$ mixing,
or the form of ${\mathcal{P}}^{(\mathrm{m})}$ of Eq.~\eqref{eq:Pmix} with $\eta-\eta'$ mixing \cite{bramon}.

The $V_{ij}$ potential can be derived from the lowest order Lagrangian of Eq.~\eqref{eq:L}.
When taking $\Phi = {\mathcal{P}}$, we get the $s$-wave $V_{ij}$ potentials without $\eta-\eta'$ mixing as
\begin{equation}
\begin{array}{ll}
V_{11}=-\dfrac{s}{2f^2},                                                       &V_{12}=-\dfrac{s-m_{\pi}^2}{\sqrt{2}f^2},\\[4mm]
V_{13}=-\dfrac{s}{4f^2},                                                       &V_{14}=-\dfrac{s}{4f^2},\\[4mm]
V_{15}=-\dfrac{m_{\pi}^2}{3\sqrt{2}f^2},                                       &V_{16}=0,\\[4mm]
V_{22}=-\dfrac{m_{\pi}^2}{2f^2},                                               &V_{23}=-\dfrac{s}{4\sqrt{2}f^2},\\[4mm]
V_{24}=-\dfrac{s}{4\sqrt{2}f^2},                                               &V_{25}=-\dfrac{m_{\pi}^2}{6f^2},\\[4mm]
V_{26}=0,                                                                      &V_{33}=-\dfrac{s}{2f^2},\\[4mm]
V_{34}=-\dfrac{s}{4f^2},                                                       &V_{35}=-\dfrac{9s-2m_{\pi}^2-6m_{\eta}^2}{12\sqrt{2}f^2},\\[4mm]
V_{36}=-\dfrac{9s-m_{\pi}^2-8m_{K}^2-3m_{\eta}^2}{12\sqrt{3}f^2},~~~~~~~~~~~~~~&V_{44}=-\dfrac{s}{2f^2},\\[5mm]
V_{45}=-\dfrac{9s-2m_{\pi}^2-6m_{\eta}^2}{12\sqrt{2}f^2},
                                                                        &V_{46}=\dfrac{9s-m_{\pi}^2-8m_{K}^2-3m_{\eta}^2}{12\sqrt{3}f^2},\\[5mm]
V_{55}=\dfrac{7m_{\pi}^2-16m_{K}^2}{18f^2},                                    &V_{56}=0,\\[4mm]
V_{66}=-\dfrac{m_{\pi}^2}{3f^2},
\end{array}
\end{equation}
with $s$ the Mandelstam variable of the scattering.

By taking $\Phi ={\mathcal{P}}^{(\mathrm{m})}$, the $s$-wave $V_{ij}$ potentials with $\eta-\eta'$ mixing are obtained,
\begin{equation}
	\begin{array}{ll}
	V_{11}=-\dfrac{s}{2f^2},                                             &V_{12}=-\dfrac{s-m_{\pi}^2}{\sqrt{2}f^2}, \\[4mm]
	V_{13}=-\dfrac{s}{4f^2},                                             &V_{14}=-\dfrac{s}{4f^2},\\[4mm]
	V_{15}=-\dfrac{\sqrt{2} m_{\pi}^2}{3f^2},                            &V_{16}=0,\\[4mm]
	V_{22}=-\dfrac{m_{\pi}^2}{2f^2},                                     &V_{23}=-\dfrac{s}{4\sqrt{2}f^2},\\[4mm]
	V_{24}=-\dfrac{s}{4\sqrt{2}f^2},                                     &V_{25}=-\dfrac{m_{\pi}^2}{3f^2},\\[4mm]
	V_{26}=0,                                                            &V_{33}=-\dfrac{s}{2f^2},\\[4mm]
	V_{34}=-\dfrac{s}{4f^2},                                             &V_{35}=-\dfrac{\sqrt{2}(3s-m_{K}^2-2m_{\eta}^2)}{9f^2},\\[4mm]
	V_{36}=-\dfrac{3s-2m_{K}^2-m_{\eta}^2}{3\sqrt{6}f^2},                &V_{44}=-\dfrac{s}{2f^2},\\[4mm]
	V_{45}=-\dfrac{\sqrt{2}(3s-m_{K}^2-2m_{\eta}^2)}{9f^2},~~~~~~~~~~~~~~&V_{46}=\dfrac{3s-2m_{K}^2-m_{\eta}^2}{3\sqrt{6}f^2},\\[4mm]
	V_{55}=-\dfrac{m_{\pi}^2+2m_{K}^2}{9f^2},                            &V_{56}=0,\\[4mm]
	V_{66}=-\dfrac{2m_{\pi}^2}{3f^2}.
	\end{array}
\end{equation}
%



\begin{thebibliography}{}

\bibitem{1us}
E.~van Beveren, T.~A.~Rijken, K.~Metzger, C.~Dullemond, G.~Rupp and J.~E.~Ribeiro,
Z. Phys. C \textbf{30}, 615-620 (1986).

\bibitem{2us}
E.~van Beveren, D.~V.~Bugg, F.~Kleefeld and G.~Rupp,
Phys. Lett. B \textbf{641}, 265-271 (2006).

\bibitem{3us}
F.~E.~Close and N.~A.~Tornqvist,
J. Phys. G \textbf{28}, R249-R267 (2002).

\bibitem{4us}
J.~R.~Pelaez,
Phys. Rev. Lett. \textbf{92}, 102001 (2004).

\bibitem{5us}
D.~Black, A.~H.~Fariborz, F.~Sannino and J.~Schechter,
Phys. Rev. D \textbf{59}, 074026 (1999).

\bibitem{6us}
F.~De Fazio and M.~R.~Pennington,
Phys. Lett. B \textbf{521}, 15-21 (2001).

\bibitem{7us}
R.~A.~Briceno, J.~J.~Dudek, R.~G.~Edwards and D.~J.~Wilson,
Phys. Rev. D \textbf{97}, no.5, 054513 (2018).

\bibitem{14us}
J.~D.~Weinstein and N.~Isgur,
Phys. Rev. D \textbf{41}, 2236 (1990).

\bibitem{8us}
S.~Weinberg,
Phys. Rev. \textbf{166}, 1568-1577 (1968).

\bibitem{9us}
J.~Gasser and H.~Leutwyler,
Annals Phys. \textbf{158}, 142 (1984).

\bibitem{10us}
J.~A.~Oller and E.~Oset,
Nucl. Phys. A \textbf{620}, 438-456 (1997);
[Erratum: Nucl. Phys. A \textbf{652}, 407-409 (1999)].

\bibitem{11us}
N.~Kaiser,
Eur. Phys. J. A \textbf{3}, 307-309 (1998).

\bibitem{12us}
M.~P.~Locher, V.~E.~Markushin and H.~Q.~Zheng,
Eur. Phys. J. C \textbf{4}, 317-326 (1998).

\bibitem{13us}
J.~Nieves and E.~Ruiz Arriola,
Nucl. Phys. A \textbf{679}, 57-117 (2000).

\bibitem{toledoliang}
S.~Sakai, W.~H.~Liang, G.~Toledo and E.~Oset,
Phys. Rev. D \textbf{101}, no.1, 014005 (2020).

\bibitem{2wang}
N.~N.~Achasov, S.~A.~Devyanin and G.~N.~Shestakov,
Phys. Lett. B \textbf{88}, 367-371 (1979).

\bibitem{3wang}
N.~N.~Achasov, S.~A.~Devyanin and G.~N.~Shestakov,
Yad. Fiz. \textbf{33}, 1337-1348 (1981); Sov. J. Nucl. Phys. \textbf{33}, 715 (1981).

\bibitem{4wang}
N.~N.~Achasov and G.~N.~Shestakov,
Phys. Rev. D \textbf{56}, 212-220 (1997).

\bibitem{5wang}
O.~Krehl, R.~Rapp and J.~Speth,
Phys. Lett. B \textbf{390}, 23-28 (1997).

\bibitem{6wang}
B.~Kerbikov and F.~Tabakin,
Phys. Rev. C \textbf{62}, 064601 (2000).

\bibitem{7wang}
F.~E.~Close and A.~Kirk,
Phys. Lett. B \textbf{489}, 24-28 (2000).

\bibitem{8wang}
A.~E.~Kudryavtsev and V.~E.~Tarasov,
JETP Lett. \textbf{72}, 410-414 (2000); Pisma Zh. Eksp. Teor. Fiz. \textbf{72}, 589-594 (2000).

\bibitem{9wang}
V.~Y.~Grishina, L.~A.~Kondratyuk, M.~Buescher, W.~Cassing and H.~Stroher,
Phys. Lett. B \textbf{521}, 217-224 (2001).

\bibitem{10wang}
F.~E.~Close and A.~Kirk,
Phys. Lett. B \textbf{515}, 13-16 (2001)

\bibitem{11wang}
A.~E.~Kudryavtsev, V.~E.~Tarasov, J.~Haidenbauer, C.~Hanhart and J.~Speth,
Phys. Rev. C \textbf{66}, 015207 (2002).

\bibitem{12wang}
L.~A.~Kondratyuk, E.~L.~Bratkovskaya, V.~Y.~Grishina, M.~Buescher, W.~Cassing and H.~Stroher,
Phys. Atom. Nucl. \textbf{66}, 152-171 (2003); Yad. Fiz. \textbf{66}, 155-174 (2003).

\bibitem{13wang}
N.~N.~Achasov and A.~V.~Kiselev,
Phys. Lett. B \textbf{534}, 83-86 (2002).

\bibitem{14wang}
N.~N.~Achasov and G.~N.~Shestakov,
Phys. Rev. Lett. \textbf{92}, 182001 (2004).

\bibitem{15wang}
V.~Y.~Grishina, L.~A.~Kondratyuk, M.~Buescher and W.~Cassing,
Eur. Phys. J. A \textbf{21}, 507-520 (2004).

\bibitem{16wang}
N.~N.~Achasov and G.~N.~Shestakov,
Phys. Rev. D \textbf{70}, 074015 (2004).

\bibitem{17wang}
J.~J.~Wu, Q.~Zhao and B.~S.~Zou,
Phys. Rev. D \textbf{75}, 114012 (2007).

\bibitem{18wang}
J.~J.~Wu and B.~S.~Zou,
Phys. Rev. D \textbf{78}, 074017 (2008).

\bibitem{19wang}
C.~Hanhart, B.~Kubis and J.~R.~Pelaez,
Phys. Rev. D \textbf{76}, 074028 (2007).

\bibitem{20wang}
J.~J.~Wu, X.~H.~Liu, Q.~Zhao and B.~S.~Zou,
Phys. Rev. Lett. \textbf{108}, 081803 (2012).

\bibitem{21wang}
F.~Aceti, W.~H.~Liang, E.~Oset, J.~J.~Wu and B.~S.~Zou,
Phys. Rev. D \textbf{86}, 114007 (2012).

\bibitem{22wang}
L.~Roca,
Phys. Rev. D \textbf{88}, 014045 (2013).

\bibitem{23wang}
V.~E.~Tarasov, W.~J.~Briscoe, W.~Gradl, A.~E.~Kudryavtsev and I.~I.~Strakovsky,
Phys. Rev. C \textbf{88}, 035207 (2013).

\bibitem{24wang}
T.~Sekihara and S.~Kumano,
Phys. Rev. D \textbf{92}, no.3, 034010 (2015).

\bibitem{25wang}
F.~Aceti, J.~M.~Dias and E.~Oset,
Eur. Phys. J. A \textbf{51}, no.4, 48 (2015).

\bibitem{vinibayar}
M.~Bayar and V.~R.~Debastiani,
Phys. Lett. B \textbf{775}, 94-99 (2017).

\bibitem{aliev}
T.~M.~Aliev and S.~Bilmis,
Eur. Phys. J. A \textbf{54}, no.9, 147 (2018).

\bibitem{chengli}
X.~D.~Cheng, H.~B.~Li, R.~M.~Wang and M.~Z.~Yang,
Phys. Rev. D \textbf{99}, no.1, 014024 (2019).


\bibitem{liangchen}
W.~H.~Liang, H.~X.~Chen, E.~Oset and E.~Wang,
Eur. Phys. J. C \textbf{79}, no.5, 411 (2019).

\bibitem{chengyu}
X.~D.~Cheng, R.~M.~Wang and Y.~G.~Xu,
Phys. Rev. D \textbf{102}, no.5, 054009 (2020).

\bibitem{beseta}
M.~Ablikim \textit{et al.} [BESIII],
Phys. Rev. Lett. \textbf{108}, 182001 (2012).

\bibitem{wuwu}
X.~G.~Wu, J.~J.~Wu, Q.~Zhao and B.~S.~Zou,
Phys. Rev. D \textbf{87}, no.1, 014023 (2013).

\bibitem{achaeta}
N.~N.~Achasov, A.~A.~Kozhevnikov and G.~N.~Shestakov,
Phys. Rev. D \textbf{92}, no.3, 036003 (2015).

\bibitem{duzhao}
M.~C.~Du and Q.~Zhao,
Phys. Rev. D \textbf{100}, no.3, 036005 (2019).

\bibitem{karplus}
R.~Karplus, C.~M.~Sommerfield and E.~H.~Wichmann,
Phys. Rev. \textbf{111}, 1187-1190 (1958).

\bibitem{landau}
L.~D.~Landau,
Nucl. Phys. \textbf{13}, no.1, 181-192 (1959).

\bibitem{abnormal}
S.~Sakai, E.~Oset and W.~H.~Liang,
Phys. Rev. D \textbf{96}, no.7, 074025 (2017).

\bibitem{isoviobdec}
W.~H.~Liang, S.~Sakai, J.~J.~Xie and E.~Oset,
Chin. Phys. C \textbf{42}, no.4, 044101 (2018).

\bibitem{daits}
L.~R.~Dai, Q.~X.~Yu and E.~Oset,
Phys. Rev. D \textbf{99}, no.1, 016021 (2019).

\bibitem{fengkunsakai}
H.~J.~Jing, S.~Sakai, F.~K.~Guo and B.~S.~Zou,
Phys. Rev. D \textbf{100}, no.11, 114010 (2019).

\bibitem{besexp}
M.~Ablikim \textit{et al.} [BESIII],
Phys. Rev. D \textbf{103}, no.9, 092004 (2021).

\bibitem{pdg}
P.A.~Zyla \textit{et al.} [Particle Data Group],
Prog. Theor. Exp. Phys. \textbf{2020}, no.8, 083C01 (2020) and 2021 update.

\bibitem{weiwang}
W.~Wang,
Phys. Lett. B \textbf{759}, 501-506 (2016).


\bibitem{review}
E.~Oset, W.~H.~Liang, M.~Bayar, J.~J.~Xie, L.~R.~Dai, M.~Albaladejo, M.~Nielsen, T.~Sekihara, F.~Navarra and L.~Roca, \textit{et al.}
Int. J. Mod. Phys. E \textbf{25}, 1630001 (2016).

\bibitem{sekisemi}
T.~Sekihara and E.~Oset,
Phys. Rev. D \textbf{92}, no.5, 054038 (2015).

\bibitem{sekinavarra}
F.~S.~Navarra, M.~Nielsen, E.~Oset and T.~Sekihara,
Phys. Rev. D \textbf{92}, no.1, 014031 (2015).

\bibitem{liangxie}
W.~H.~Liang, E.~Oset and Z.~S.~Xie,
Phys. Rev. D \textbf{95}, no.1, 014015 (2017).

\bibitem{pavaoliang}
R.~P.~Pavao, W.~H.~Liang, J.~Nieves and E.~Oset,
Eur. Phys. J. C \textbf{77}, no.4, 265 (2017).

\bibitem{bramon}
A.~Bramon, A.~Grau and G.~Pancheri,
Phys. Lett. B \textbf{283}, 416-420 (1992).

\bibitem{LiangOset}
W.~H.~Liang and E.~Oset,
Phys. Lett. B \textbf{737}, 70-74 (2014).

\bibitem{DaiXie}
J.~J.~Xie, L.~R.~Dai and E.~Oset,
Phys. Lett. B \textbf{742}, 363-369 (2015).

\bibitem{liangchic1}
W.~H.~Liang, J.~J.~Xie and E.~Oset,
Eur. Phys. J. C \textbf{76}, no.12, 700 (2016).

\bibitem{kornicer}
M.~Ablikim \textit{et al.} [BESIII],
Phys. Rev. D \textbf{95}, no.3, 032002 (2017).

\bibitem{hietala}
J.~Hietala, D.~Cronin-Hennessy, T.~Pedlar and I.~Shipsey,
Phys. Rev. D \textbf{92}, no.1, 012009 (2015).


\end{thebibliography}
\end{document}